\documentclass[twocolumn,showpacs,preprintnumbers,amsmath,amssymb,superscriptaddress]{revtex4}

\usepackage{graphicx}
\usepackage{dcolumn}
\usepackage{bm}
\usepackage{longtable}

\begin{document}


\title{Exact mean first-passage time on the T-graph}

\author{E. Agliari}
\affiliation{Dipartimento di Fisica, Universit\`a degli Studi di
Parma, viale Usberti 7/A, 43100 Parma, Italy}

\date{\today}

\begin{abstract}
We consider a simple random walk on the T-fractal and we calculate
the exact mean time $\tau^g$ to first reach the central node
$i_0$. The mean is performed over the set of possible walks from a
given origin and over the set of starting points uniformly
distributed  throughout the sites of the graph, except $i_0$. By
means of analytic techniques based on decimation procedures, we
find the explicit expression for $\tau^g$ as a function of the
generation $g$ and of the volume $V$ of the underlying fractal.
Our results agree with the asymptotic ones already known for
diffusion on the T-fractal and, more generally, they are
consistent with the standard laws describing diffusion on
low-dimensional structures.

\end{abstract}

\pacs{05.40.Fb} \maketitle

\section{\label{sec:intro}Introduction}
Many problems in physics and chemistry are related to random walks
on fractal structures \cite{havlin1,balakrishnan,hughes}. The main
reason is that such structures are able to mimic the inhomogeneity
and scale-invariance typical of disordered materials. An important
class of fractal structures is given by the so called exactly
decimable fractals which include deterministic finitely ramified
fractals such as the Sierpinski gasket and the T-graph. These
structures, being amenable to renormalization procedures, allow
exact analytic calculations \cite{machta,van1,burioni}.

In general, the lack of translational invariance implies
significant corrections to the standard laws describing diffusion
on regular lattices. Indeed, in the fields of reaction-diffusion
and transport theory, a question of longstanding interest concerns
the interplay between spatial extent and system dimensionality in
affecting the reaction kinetics and the transport efficiency
\cite{kozak2}.\\
A fundamental quantity characterizing diffusion is the mean
first-passage time (MFPT), i.e. the expected time for a random
walker, starting with equal probability at any site $i \neq i_0$,
to first reach a given site $i_0$. This problem was first set up
by Montroll \cite{montroll} in the case of regular structures and
later extended to more complex substrates
\cite{matan,kahng,van,kozak,baronchelli,redner,condamin}. Notice
that this definition of MFPT involves a double average: the first
one is over all the walks from a given origin $i$, then you must
average over a uniform distribution of initial sites, whose
support is the
whole set of graph sites, except $i_0$.\\
The mean first-passage time is also intrinsically related to a
number of different problems \cite{redner}. In the context of
reaction-diffusion processes it represents the mean time to react
for a particle diffusing in the presence of an active site located
in $i_0$ \cite{blumen,agliari2,montroll,weiss}, which is sometimes
referred to as target problem. Not only: the mean first-passage
time defined above also describes the asymptotic behaviour of the
average time for two diffusive particles to first encounter
\cite{oshanin}.

The MFPT has been previously studied on different kinds of
structures and several analytical results have been found. Most of
them consists of scaling relations and asymptotic behaviours
\cite{kahng,van,baronchelli}, while a very few exact results are
known \cite{kozak,redner}. Exact solutions on finite structures
are especially longed for since they prove useful for a deeper
comprehension of theoretical models and for checking approximate
solutions or numerical simulations.

\begin{figure}[tb] \begin{center}
\includegraphics[width=.36\textwidth]{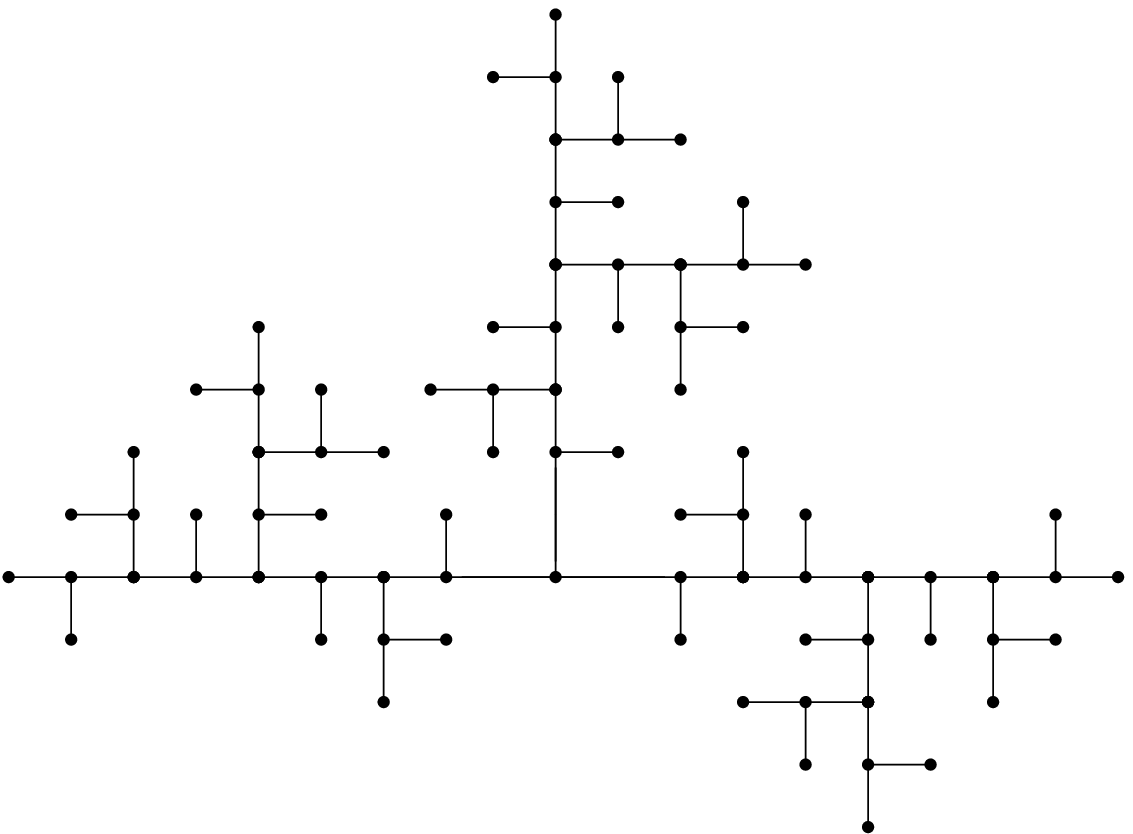}
\includegraphics[width=.12\textwidth]{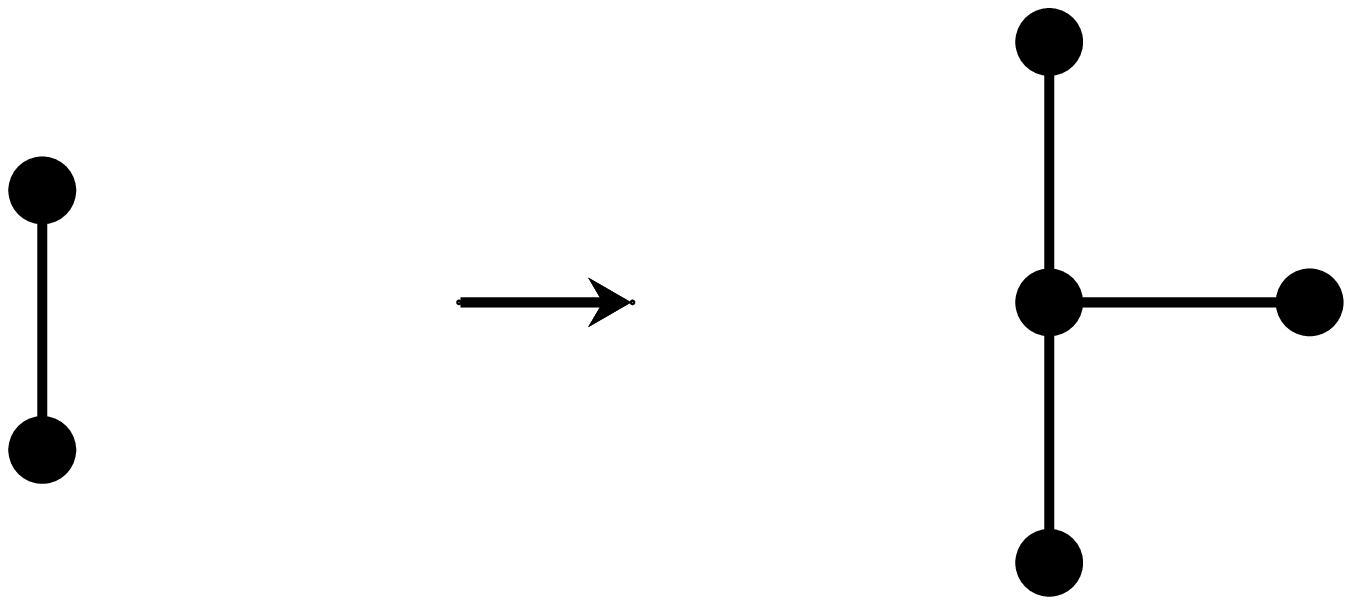}
\caption{\label{fig:tfractal} T-fractal of generation 4:
$V=3^4+1$. The next generation is obtained by performing the
operation illustrated in the bottom on each bond.}\end{center}
\end{figure}

Here, we derive the exact mean first-passage time for a random
walker on the T-fractal, following the decimation procedure
recently introduced by Kozak et al. and applied to the Sierpinski
gasket \cite{kozak}. In particular, we assume a simple random
walker (RW) in the presence of a perfect trap fixed at the central
site and we calculate the mean-walk length before absorption. The
closed-form expression we obtain for the latter, is akin to the
one in \cite{kozak} and consistent with known asymptotic results
\cite{oshanin,matan,kahng}.\\
Though they are both deterministic fractals, the Sierpinski gasket
and the T-fractal display significant differences: while the
former models self-similar structures endowed with closed loops,
the T-fractal is representative for tree-like structures and
bundled structures in general
\cite{burioni1,burioni2,maritan,knezevic,kahng,matan}. For this
reason it is worth extending and comparing the related results.

The paper is organized as follows. In Sec.~\ref{sec:theory} we
describe the main features of the T-graph and we resume the
analytic background underlying the analytic solution; in
Sec.~\ref{sec:Res} we describe the decimation procedure applied
and we obtain the exact formula for the mean first-passage time as
a function of both the generation and the volume of the structure;
finally Sec.~\ref{sec:Conclusions} includes conclusions and
comments.

\section{\label{sec:theory}Theory}
\subsection{\label{sec:Decimable}Exactly Decimable Fractals}

A generic graph $\mathcal{G}$ is mathematically specified by the
pair $\{\Lambda, \Gamma \}$ consisting of a non-empty, countable
set of points, $\Lambda$ joined pairwise by a set of links
$\Gamma$. The cardinality of $\Lambda$ is given by $|\Lambda| = V$
representing the number of sites making up the graph, i.e. its
volume. From an algebraic point of view, a graph
$\mathcal{G}=\{\Lambda, \Gamma \}$ is completely described by its
adjacency matrix $A$. Every entry of this off-diagonal, symmetric
matrix corresponds to a pair of sites, and it equals one if and
only if this couple is joined by a link, otherwise it is zero. The
number of nearest-neighbours of the generic site $i$, referred to
as coordination number, can be recovered as a sum of adjacency
matrix elements: $z_i = \sum_{j \in \Lambda} A_{ij}$. These are
used to build up the diagonal matrix $Z_{ij}=z_i \delta_{ij}$
\cite{burioni}.\\
A very special class of graphs is given by the so called
\textit{exactly decimable fractals} which are geometrically
invariant under site decimation. In general, all deterministic,
finitely ramified fractals are exactly decimable. The solution of
both the random walk and harmonic oscillations problems can be
obtained by standard renormalization group calculations based on
real space decimation procedures \cite{machta,van1,burioni}. The
Sierpinski gasket, the T-graph (Fig.~\ref{fig:tfractal}), the
branched Koch curves are examples of exactly decimable fractals,
which accounts for their popularity. Notice that such structures
are characterized by strong restrictions on their topology which
can give rise to properties far from holding for all fractals
\cite{burioni1}.

%
%
Here, we consider a T-fractal which is iteratively constructed by
performing the operation illustrated in Fig.~\ref{fig:tfractal} on
each link. The number of iterations is called the generation $g$
of the fractal. At the $g$-th generation the cardinality of the
set of nodes $V(g)$, hereafter called volume, is given by: $V(g)
\equiv |\Lambda_g|=3^{g}+1$. The T-fractal has fractal dimension
$d_f=\frac{\log3}{\log2}\approx1.584$ and spectral dimension
$\tilde{d}=\frac{\log 9}{\log 6}\approx1.226.$ We recall that the
former gives the dependence of the volume of the system on its
linear size $L$:
$$
V(g)\sim \left[ L(g) \right] ^{d_f} = (2^g)^{d_f},
$$
while the latter governs (among other phenomena) the long-time
properties of diffusion on the graph. Indeed, if we consider a
random walker starting from a given site $i$ of the graph, the
probability $P_{ii}(t)$ of returning back to the starting point at
time $t$, at long times, follows the law
$$
P_{ii}(t)\sim t^{-\tilde{d}/2}.
$$
When $\tilde{d} < 2$ the random walker is said to perform a
``compact exploration'' of the space \cite{degennes} since the
fractal dimension of its trajectory is greater than the dimension
$d_f$ of the substrate.

It is worth underlining that the T-fractal is irregular, that is
the coordination number is site dependent. We can distinguish
among ``internal site'' with coordination number $z=3$ and
``external site'' with $z=1$. We call $\Lambda_{int}$ and
$\Lambda_{ext}$ the set of internal and external sites
respectively. Obviously $\Lambda^g \equiv \Lambda^g_{int} \cup
\Lambda^g_{ext}$.

\begin{figure}[tb] \begin{center}
\includegraphics[width=.52\textwidth]{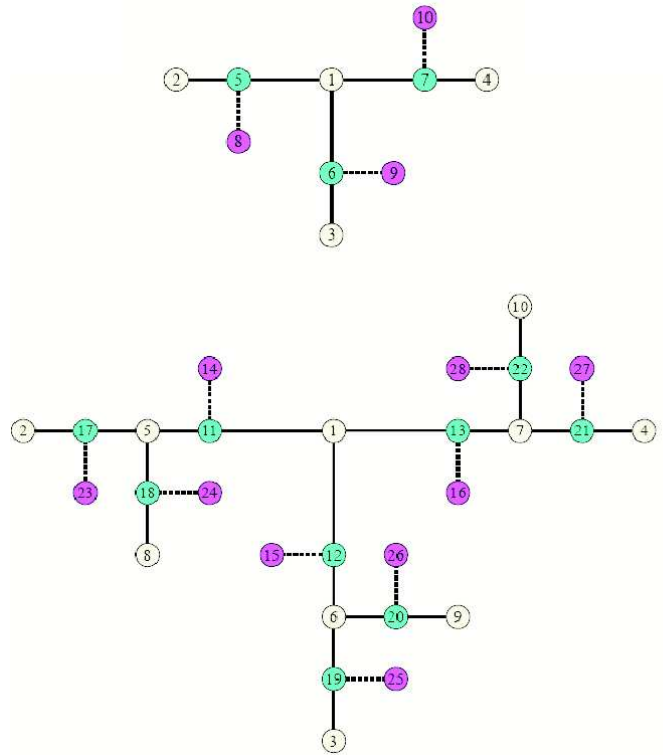}
\caption{\label{fig:color} (Color on line) T-graph of generation
$2$ (top) and $3$ (bottom) with volume $V=3^2+1$ and $V=3^3+1$,
respectively. Each site has been labelled according to the
procedure described in the text. Dotted links belong to
$\bar{\Gamma}^g$ and colored sites to $\bar{\Lambda}^g$; more
precisely, pink sites are in $\bar{\Lambda}_{ext}^g$ and blue
sites in $\bar{\Lambda}_{int}^g$ }.
\end{center}
\end{figure}

Figure \ref{fig:color} shows generations $g=2,3$ and the labelling
method adopted, which will be useful in the next section. In each
new generation, we label only the new sites while old sites keep
their own labels. Hence, at generation $g+1$ we name new sites
progressively, starting from $V(g)+1$. The first sites to be
labelled are the innermost, the last ones are those farthest from
the central site $i_0=1$. At each generation we can distinguish
sets of equivalent sites which are labelled anti-clock wise. Due
to the symmetry of the T-graph, the cardinality of such sets is
always a multiple of $3$; for example we have $\{2,3,4,8,9,10
\}_{g=2}, \{14,15,16 \}_{g=3}$.

Before proceeding further, let us resume some facts concerning
unbiased diffusion on a T-fractal of generation $g$ represented by
the adjacency matrix $A_g$ (henceforth we will omit the subscript
$g$). We consider a simple RW, starting, at $t=0$, from site $i$;
at each time step (taken to be unity) the particle jumps with
equal probability to any of its nearest-neighbour sites. Being
$P_{ji}(t)$ the probability of going from $i$ to $j$ in $t$ steps,
the following Master equation holds:
\begin{equation}\label{eq:master}
P_{ji}(t+1) = \sum_{k=1}^{V} (Z^{-1}A)_{jk} P_{ki}(t),
\end{equation}
which states that at each time step, the jumping probability from
an internal site is $\frac{1}{3}$. From the previous equation it
follows that $P_{ji}(t)=[(Z^{-1}A)^t]_{ji}$. It is also easy to
verify that the Markov chain representing such a random walk is
ergodic and the particle will visit all sites with probability
$1$, independently of the origin $i$. Consequently, the walker
will reach any site with probability $1$, in a time possibly
diverging when $g \rightarrow \infty$.

\subsection{\label{sec:MeanTime}Mean Time to Absorption}

Let us consider a perfectly absorbing trap, fixed on the central
site (labelled with index $i_0 = 1$) of the T-graph. Our aim is to
obtain, through a decimation procedure, an exact, closed-form
solution for the average time to absorption, where the average is
meant both over all the possible walks starting from the same
origin $i$ and over all sites $i \neq i_0$ taken as origin of the
walk.\\
The special choice we made for the trap location makes the
decimation procedure easier to be applied as we can identify the
site $i_0$ since the first generation.

We now introduce $\tau_{i,q}^{g}$ to be the $q$-th ($q=0,1,...$)
moment of the trapping time for a walk starting from $i$ on the
$g$-th generation of the graph. Obviously, regardless of $g$,
$\tau_{0,q}^{g}=0$ and $\tau_{i,0}^{g}=1$. The latter is the
zero-th moment of the distribution of the time to absorption given
$i$ as origin and it is unitary because the walker will be trapped
with probability $1$, whatever its origin.

The starting point for our analytic treatment is the discrete
differential equation introduced in \cite{kozak}:
\begin{equation}\label{eq:koz}
- \sum_{j=2}^{V(g)} \Delta_{ij} \tau_{j,\,q+1}^{g} = (q+1)
\tau_{i,\,q}^{g},
\end{equation}
where $\Delta = A Z^{-1} - \mathrm{I}$ is a normalized version of
the discrete Laplacian whose first row and column (corresponding
to the trap site) have been removed. We recall that $\Delta$ is a
nonsingular matrix and each row has sum zero, apart from those
corresponding to sites $V(g-1)+1$, $V(g-1)+2$ and $V(g-1)+3$, i.e.
the three nearest-neighbours of the trap, for which the sum is $-
\frac{1}{3}$.
As discussed in \cite{kozak2}, Eq.~\ref{eq:koz} can be generalized to the case of two or
more particles simultaneously diffusing.

In the following, we just focus on the set of first moments
$\tau_{j,1}^{g}$, for which Eq.~(\ref{eq:koz}) simplifies into
$$
- \sum_{j} \Delta_{ij} \tau_{j,1}^g = \tau_{i,0}^g = 1.
$$
Henceforth we can drop the index corresponding to $q$ without
ambiguity:
$$
- \sum_{j} \Delta_{ij} \tau_{j}^g = 1.
$$
Now, we implement the average over the starting site $i$, chosen
according to a uniform distribution in $\Lambda^g \setminus \{ i_0
\}$:
\begin{equation}\label{eq:tau_av}
\tau^{g} = \frac{1}{V(g)-1} \sum_{i=2}^{V(g)} \tau_i^{g} =
\frac{1}{V(g)-1} \sum_{i=2}^{V(g)}\sum_{j=2}^{V(g)}
(-\Delta^{-1})_{ij}.
\end{equation}
In the next Section we derive some recurrence relations which
allow to simplify the previous equation.

Equation \ref{eq:tau_av} can be very easily interpreted if we look
at the random walk as a Markov chain. Indeed, $-\Delta^{-1}$ is
just the fundamental matrix for the process, whose entry $i,j$
represents, by definition, the expected number of times that the
process is in the transient state $j$, being started in the
transient state $i$.

Finally, notice that, due to the symmetry and the absence of loops
characterizing the structure under consideration, the mean time to
absorption found in this case just corresponds to the mean time to
reach either site $2,3,4$ on a T-fractal of generation $g-1$.
Otherwise stated, if we call $\pi^g$ the mean time to first reach
site $2$ (or, analogously, $3,4$), then $\pi^g=\tau^{g+1}$.

\section{\label{sec:Res} Decimation procedure}

The number of terms to sum up in Eq.~\ref{eq:tau_av} grows
exponentially with $g$, hence, a direct calculation of $\tau_i^g$
and $\tau^g$ can be accomplished straightforwardly only for the
very first generations (see Tab.~\ref{tab:values_2} and
\ref{tab:values_1}). Such data allow to get some recurrence
relations useful for the derivation of the final formula. First of
all, notice that for a given site $i$, we have $\tau_i^{g+1}= 6
\tau_i^{g}$: in each generation the chemical distance from $i$ to
the trap doubles while the mean time to first reach the trap
increases by a factor $6$. This exact scaling follows from the
symmetry and decimability of the graph and it is consistent with
the random walk dimension on the T-fractal: $d_w =
\frac{2d_f}{\tilde{d}}=\frac{\log 6}{\log 2}$ \cite{kozak}.
\begin{table*}
\caption{Mean time to absorption $\tau^g_i$ for a random walker
starting from a given site $i$. For these values the average is
only performed over all possible random walks sharing the same
origin. Due to the symmetry of the T-fractal we can distinguish
sets of equivalent sites such that, if taken as origin of the
walk, they provide the same mean time $\tau^g_i$. Notice that, for
the farthest sites from the trap, having chemical distance
$2^{g-1}$, the average absorption time is $6^{g-1}$, consistently
with the random walk dimension $d_w$.}
\begin{ruledtabular}
\begin{tabular}{cccccccccc}
$g \backslash i$ & (2, 3, 4) & (5, 6, 7) & (8, 9, 10) & (11, 12,
13) & (14, 15, 16) & (17, 18, 19, 20, 21, 22) & (23, 24, 25, 26,
27, 28) &
(29,30,31) & (32, 33, 34)\\
\hline
1 &   1 &     - &    - &   - &    - &    - &    - &   - &    - \\
2 &   6 &     5 &    6 &   - &    - &    - &    - &   - &    - \\
3 &  36 &    30 &   36 &  17 &   18 &   35 &   36 &   - &    - \\
4 & 216 &   180 &  216 & 102 &  108 &  210 &  216 &  53 &   54 \\
\end{tabular}
\end{ruledtabular}
\label{tab:values_2}
\end{table*}

Furthermore, at each generation $g$ we insert on the existing
fractal some new links and some new vertices. We call such sets
$\bar{\Lambda}^{g} = \Lambda^{g} \setminus \Lambda^{g-1}$ and
$\bar{\Gamma}^{g}=\Gamma^g \setminus \Gamma^{g-1}$, respectively.
It is easy to see that $|\bar{\Lambda}^{g}|=V(g)-V(g-1)=2 \cdot
3^{g-1}$ and $2 \cdot |\bar{\Gamma}^{g}| = |\bar{\Lambda}^{g}|$.
Moreover, for each new link added we have a new couple of
connected vertices $j_{ext}$ and $j_{int}$, belonging to
$\bar{\Lambda}_{ext}^{g}$ and $\bar{\Lambda}_{int}^{g}$, whose
coordination numbers are $z_{j_{ext}}=1$ and $z_{j_{int}}=3$,
respectively and:
$$
\bar{\Lambda}^{g} = \bar{\Lambda}_{ext}^{g} \cup
\bar{\Lambda}_{int}^{g},
$$
\begin{equation}\label{eq:relutile1}
\frac{1}{2} \bar{\Lambda}^{g} = \bar{\Lambda}_{ext}^{g} =
\bar{\Lambda}_{int}^{g}.
\end{equation}
For example, $\bar{\Lambda}_{ext}^{3}=\{14, 15, 16, 23,24, 25, 26,
27, 28 \}$, as shown in Fig.~\ref{fig:color}. Now, it is easy to
see that
\begin{equation}\label{eq:relutile2}
\tau_{i_{ext}}^{g}=\tau_{i_{int}}^{g} +1
\end{equation}
since a RW starting from $i_{ext}$ is necessarily on $i_{int}$ at
time $t=1$.
These facts hold regardless of the generation $g$ and for any
connected couple chosen from $\bar{\Lambda}^{g}$. Thus we can
write
\begin{equation}\label{eq:recurrence0}
\sum_{i=2}^{V(g)} \tau_i^{g}= 6 \sum_{i \in \Lambda^{g-1}}
\tau_i^{g-1} + \sum_{i \in \bar{\Lambda}^{g}} \tau_i^{g}.
\end{equation}
In the last sum we can highlight the contribution from external
and internal sites and, exploiting Eqs.~\ref{eq:relutile1} and
\ref{eq:relutile2}:
\begin{equation}\label{eq:recurrence1}
\sum_{i \in \bar{\Lambda}^{g}} \tau_i^{g} = \sum_{i \in
\bar{\Lambda}_{ext}^{g}} \tau_i^{g} + \sum_{i \in
\bar{\Lambda}_{int}^{g}} \tau_i^{g} =  2 \sum_{i \in
\bar{\Lambda}_{int}^{g}} \tau_i^{g} + |\bar{\Lambda}_{int}^{g}|.
\end{equation}
%
Let us now focus on the sum appearing in the right-most side and
estimate it in the case $g=3$ depicted in Fig.~\ref{fig:color}.
The mean time to absorption for a RW starting from site $5$ can be
expressed as:
\begin{equation}\label{eq:recurrence2}
\begin{aligned}
\tau^3_{5} & = 1+ (P_{11,5} \tau_{11}^{3} + P_{17,5} \tau_{17}^{3}
+ P_{18,5} \tau_{18}^{3}) =\\
& = \frac{(1+\tau_{11}^3) + (1+\tau_{17}^3) + (1+\tau_{18}^3)}{3}=\\
& = \frac{\tau_{14}^3 + \tau_{23}^3 + \tau_{24}^3}{3},
\end{aligned}
\end{equation}
where $P_{ki}$ represents the transition probability from state
$i$ to state $k$. Hence, $\tau_5^3$ is just the average of the
absorption times from $11,17,18$, which mirrors the barycentric
position of $5$ with respect to the latters. Also notice that site
$5$ belongs to $\bar{\Lambda}_{int}^{g-1}$. Since these facts hold
for any analogous subtree of generation $2$, we are allowed to
write:
\begin{equation}\label{eq:recurrence3}
\sum_{i \in \bar{\Lambda}_{ext}^{g}} \tau_{i}^g = 3 \cdot 6
\sum_{i \in \bar{\Lambda}_{int}^{g-1}} \tau_i^{g-1}.
\end{equation}
From Eqs.~\ref{eq:recurrence1} and ~\ref{eq:recurrence3}:
\begin{equation}\label{eq:recurrence4}
\sum_{i \in \bar{\Lambda}_{int}^{g}} \tau_{i}^g +3^{g-1} = 18
\sum_{i \in \bar{\Lambda}_{int}^{g-1}} \tau_{i}^{g-1},
\end{equation}
and solving this recurrence relation we obtain:
\begin{equation}\label{eq:recurrence5}
\sum_{i \in \bar{\Lambda}_{int}^{g}} \tau_{i}^g =\frac{3^{g-1}}{5}
(1+4\cdot 6^{g-1}).
\end{equation}
By plugging the last expression into Eq.~(\ref{eq:recurrence0}) we
get
$$
\begin{aligned}
(V(g)-1) \tau^{g} = & 6(V(g-1)-1) \tau^{g-1} + \\ & + \frac{2}{5}
\cdot 3^{g-1} (1+4\cdot 6^{g-1}) + 3^{g-1} \end{aligned}
$$
and, dividing by $V(g)-1=3^g$:
\begin{equation}\label{eq:abstime_t}
\tau^{g} = 2 \tau^{g-1} + \frac{1}{15} (7+ 8 \cdot 6^{g-1}).
\end{equation}
The last expression is, again, a recursive equation, whose
solution provides the exact time to absorption:
\begin{equation}\label{eq:recurrence6}
\tau^{g} = \frac{1}{15} (-7+ 5 \cdot 2^{g} + 2 \cdot 6^{g}).
\end{equation}
A numerical check of this formula can be attained by comparing
$\tau^g$, $ 1\leq g \leq 6$ with data obtained by direct
calculation and reported in Tab.~\ref{tab:values_1}: the agreement
is perfect.

\begin{table}
\caption{Mean first-time $\tau_g$ obtained by direct calculation
from Eq.~\ref{eq:tau_av}; since this implies a sum over a number
of terms exponentially increasing with $g$, only small generations
have been considered.} \label{tab:values_1}
\begin{ruledtabular}
\begin{tabular}{ccc}
g & V(g) & $\tau ^{g}$\\
\hline
1&    4&    1 \\
2&   10&    51/9 \\
3&   28&    837/27 \\
4&   82&    14391/81 \\
5&  244&    254421/243 \\
6&  730&    4550175/729 \\
\end{tabular}
\end{ruledtabular}
\end{table}

It is also possible to obtain an expression for $\tau^{g}$ as a
function of the volume $V(g)=3^{g} + 1$. In fact, recalling the
spectral dimension for the T-graph, we can write $2^{g} = (V(g) -
1)^{2/\tilde{d}-1}$, and
\begin{equation}\label{eq:abstime_t2}
\tau^{g} = \frac{1}{15} \left[ 5 (V(g) -1)^{2/\tilde{d}-1} + 2
(V(g) - 1)^{2/\tilde{d}} -7\right].
\end{equation}
Notice that the last expression gives the exact, explicit
dependence of $\tau^g$ on $V(g)$. In the asymptotic limit, as
$V(g)$ diverges,
\begin{equation}\label{eq:abstime_t3}
\tau^{g} \rightarrow V(g)^{2/\tilde{d}}.
\end{equation}
This result is consistent with the leading behaviour of $\tau^g$
on the T-fractal discussed in \cite{kahng} and, more generally, to
the leading behaviour of the trapping time on low dimensional
($\tilde{d} < 2$) structures. In fact, as already remarked, the
MFPT represents the mean trapping time $\tau_{\mathrm{trap}}$ for
a diffusing particle in the presence of a fixed perfect trap (or,
symmetrically, the trapping time for an immobile target in the
presence of a diffusive trap). Being $P_s(t)$ the survival
probability at the $t$-th step $P_s(t)$, i.e. the probability that
the RW has not yet reached the trap site, then \cite{havlin1}:
$$
\tau_{\mathrm{trap}} = \int_{0}^{\infty}  -\frac{\partial
P_s(t)}{\partial t} \cdot t \,\, dt.
$$
The asymptotic expression for the survival probability is
\cite{blumen1}:
\begin{equation}\label{eq:survival1}
P_s(t) = \exp[-\frac{\phi(t)}{V}],
\end{equation}
with
\begin{equation}\label{eq:survival}
\phi(t) \sim \left \{
\begin{array}{cl}
t^{\tilde{d}/2}, & \tilde{d}<2 \\
\frac{t}{\log t}, & \tilde{d}=2 \\
t, & \tilde{d}>2 \end{array} \right.
\end{equation}
One therefore expects that, for low dimensional structures,
$\tau_{\mathrm{trap}} \sim V^{2/\tilde{d}}$. Our result and the
one in \cite{kozak} establish this relationship rigorously.

%

As can be shown by scaling arguments, on long times, the factor
$6^{g} \sim V(g)^{2/\tilde{d}}$, which is the leading term in
Eqs.~\ref{eq:recurrence6}, \ref{eq:abstime_t2}, is involved in all
the dynamical properties of diffusion on the T-fractal. For
example, the characteristic time in the exponential decay of the
survival probability increases with the generation of the tree as
$6^{g}$ \cite{kahng}.

\begin{figure}[tb] \begin{center}
\includegraphics[width=.5\textwidth]{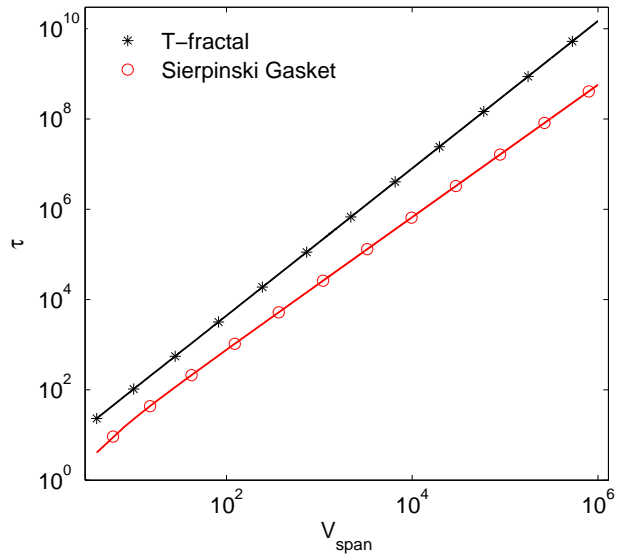}
\caption{\label{fig:compare} (Color on line) Mean first-passage
time for a simple random walker moving on a T-fractal (black
stars) and on a Sierpinski gasket (red circles) as a function of
the volume $V_{span}$ (the volume the RW can actually span before
being trapped), according to exact analytic solutions.}
\end{center}
\end{figure}

Equation \ref{eq:abstime_t2} is also consistent with a recent
result obtained by Condamin et al. \cite{condamin} who found the
asymptotic (large $V$) expression for the average time $\tau_r$
taken by a RW on a generic scale-invariant structure to first
reach a trap distant $r$ from the starting point.
%
%
In particular, for compact exploration ($d_w \geq d_f$), $\tau_r
\sim V \cdot r^{d_w - d_f}$. Thus, if we fix the trap on a
particular site $i_0$, we can obtain an estimate for the MFPT, by
simply averaging over all possible distances $r$ from $i_0$. For
the T-fractal considered here, $i_0 = 1$ and we can write:
$$
\tau^g = \frac{\sum_{r=0}^{L(g-1)} \tau_r n(r)}{V(g)},
$$
where $n(r)$ is the number of sites distant $r$ from $i_0$ and
$L(g-1)$ is the largest distance from the central site. Under the
above mentioned assumption of large volume, we can adopt a
continuous picture and $n(r) \sim 3\cdot r^{d_f-1}$; by
integrating the previous expression, we get $\tau^{g} \sim
V(g)^{2/\tilde{d}}$, as expected.

Finally, in Fig.~\ref{fig:compare} we compare
Eq.~\ref{eq:recurrence5} with the analogous formula found for the
Sierpinski gasket ($\tilde{d}=\frac{\log 9}{\log 5}$) in
\cite{kozak}:
%
$$
\tau^g=\frac{2V(g)-3}{V(g)-1} \left [
\frac{(2V(g)-3)^{\frac{2}{\tilde{d}}}}{6}+
\frac{2(2V(g)-3)^{\frac{2}{\tilde{d}}-1}}{5}-\frac{1}{6} \right ].
$$
%
In the asymptotic limit, $\tau^g$ diverges faster for the
T-fractal. This can be trivially drawn algebraically while, from a
topological point of view, it evidences the role of loops in
reducing the average distance between two random sites, making the
diffusive particles survive shorter.

\section{\label{sec:Conclusions} Conclusions}

In this work we study the mean first-passage time $\tau^g$ for a
random walker on a T-fractal. The latter, being exactly decimable,
allows the use of the powerful technique of exact renormalization.
We find an exact, closed-form solution for $\tau^g$ as a function
of either the generation $g$ and the volume $V(g)$. The leading
term of $\tau^g$ is consistent with known asymptotic results. It
should be underlined that an exact solution on a finite system is
generally very useful in order to understand more quantitatively
the asymptotic limit.

Our findings are interesting also in the light of a recent result
concerning the survival probability for the trapping problem $A +
B \rightarrow B$, with both species, $A$ and $B$, diffusing
\cite{oshanin}. There, it was shown that on low-dimensional
structures ($d<2$) the survival probability for an A particle
asymptotically does not depend on its diffusivity constant $D_A$.
Otherwise stated, at long times, the target problem and the
trapping problem provides the same results. Hence, exact results
concerning the target problem, also provide the correct asymptotic
behaviour for the trapping problem with diffusive traps.

\vspace{0.8cm}

The author is grateful to D. Cassi and R. Burioni for useful
discussions and comments.

\end{document}